# Active Passivation Tunes Hotspot Locations in GaN Transistors with In-Situ Thermal-Mechanical Visualization

Yicheng Wei[1,+], Sihang Liu[2,+], Zimu Jiang[1], jinquan Zhang[1], Zifeng Huang[2], Han Yang[3], Yang He[2], Jin Wei[2,*], Zhe Cheng[1,2,*]

[1] School of Software & Microelectronics, Peking University, Beijing 100871, China

[2] School of Integrated Circuits and Beijing Advanced Innovation Center for Integrated Circuits, Peking University, Beijing 100871, China

[3] School of physics, Peking University, Beijing 100871, China

[+]These authors contributed equally

[*]Authors to whom correspondence should be addressed: jin.wei@pku.edu.cn; zhe.cheng@pku.edu.cn

**Abstract**

Efficient thermal dissipation has become critical in emerging electronic devices. However, most existing studies have primarily focused on engineering heat dissipation pathways, largely overlooking the intrinsic behavior of the heat source itself. We demonstrate an active passivation (AP) technology that proactively tunes hotspot locations in GaN transistors. By adjusting the active passivation layer length, the hotspot is shifted from the gate edge to the drain-side AP edge, establishing a clear one-to-one spatial correlation. In-situ thermal-mechanical visualization via micro-Raman thermography, combined with multi-physics electro-thermal-mechanical simulations, directly captures the spatial redistribution of both temperature and thermal stress profiles. Electrical analysis confirms that this hotspot migration is driven by the spatial shift of the peak electric field and localized Joule heating. This proactive heat-source tuning strategy provides critical design guidelines for power electronics.

## Introduction

Gallium nitride (GaN) high electron mobility transistors (HEMTs) are the workhorse of next-generation high-power systems [1,2] due to their exceptional physical properties, including a wide bandgap, a high breakdown field, and a high electron saturation velocity [3,4]. However, as operating power densities continue to escalate, severe localized Joule heating has become the primary performance bottleneck [5,6]. Under high-voltage operation, the intense electric field peaks at the drain-side edge of gate contact [7,8], resulting in a highly localized hotspot [9–11]. This localized thermal spike not only drastically degrades electron mobility, but also induces thermal tensile or compressive stresses due to thermal expansion mismatches, severely degrading device performance and long-term reliability.

To mitigate the thermal bottleneck, conventional device-level and die-level thermal management strategies have been formulated by focusing on engineering external heat dissipation pathways and minimizing overall thermal resistance [12]. Common approaches include replacing low-thermal-conductivity substrates (e.g., sapphire [13]) with high-thermal-conductivity alternatives such as SiC [12,14] and diamond [15], implementing flip-chip bonding [16,17], integrating localized heat spreaders adjacent to the active channel [12,18] as well as exploring advanced packaging and alternative cooling configurations [19,20]. Despite its demonstrated efficacy in reducing thermal resistance, this paradigm overlooks the intrinsic coupling between heat generation and device physics, wherein the heat source evolves in response to carrier transport, electric field distribution, and device architecture. As a result, thermal design is performed without explicitly considering the evolution of the heat source itself, which limits the ability to fully capture device-level thermal behaviors.

To bridge this gap, proactive structural engineering via active passivation (AP) technology offers a strategic pathway to regulate internal electric field dynamics. In an active-passivation p-GaN gate HEMT (AP-HEMT), a thin p-GaN active passivation layer is deliberately extended from the gate edge towards the drain [21], which introduces an intentional, designable mechanism for field regulation. The configuration effectively shifts the channel pinch-off point – and consequently, the peak electric field – from the gate edge to the drain-side end of the active passivation layer [22]. This deterministic migration of the peak electric field fundamentally reshapes the spatial distribution of localized Joule heating, enabling the systematic engineering of temperature and thermal stress profiles through device layout design.

In fact, while various electric field engineering techniques (e.g., field plates [23,24] and overlapping gate [25]) have been developed in GaN devices, their design goals are almost exclusively restricted to optimizing electrical characteristics, such as boosting breakdown voltage, suppressing current collapse, and reducing ON-resistance [26]. Although a few studies have numerically simulated the regulation of heat sources by electric fields [27], direct in-situ experimental visualization of such spatial hotspot evolution and thermal-mechanical profiles has yet to be demonstrated.

In this work, we demonstrate proactive hotspot tuning in GaN transistors via AP technology, combined with multi-physics analysis and in-situ visualization. First, electrical simulations were performed, revealing that the active passivation layer shifts the peak electric field and localized Joule heating from the gate edge toward the drain-side end of the layer. Subsequently, utilizing high-resolution micro-Raman mapping, we directly captured the spatial redistribution of both temperature and thermal stress profiles under various operating power levels. By systematically evaluating a series of AP-HEMTs with varying active passivation layer lengths ($L_{AP}$), a clear one-to-one spatial correlation between the hotspot location and $L_{AP}$ is experimentally established. Finally, coupled thermal-mechanical simulations were conducted to further validate the experimental observations. This paper provides a comprehensive electro-thermal-mechanical co-optimization strategy utilizing AP technology for engineering spatial distributions of electric fields, heat generation, and thermal stress in GaN HEMTs. More broadly, it establishes a general design strategy for actively modulating coupled multi-physics responses in high-power electronics through device-level structural engineering.

**Sample structure**

Both conventional p-GaN gate HEMTs (Conv-HEMTs) and active-passivation p-GaN gate HEMTs (AP-HEMTs), with varying gate-to-drain distance ($L_{GD}$), were fabricated on a commercial enhancement-mode (E-mode) AlGaN/GaN epitaxial wafer. The only structural distinction between the two devices is the integration of an active passivation layer in the AP-HEMT, as shown in the comparison of Figure 1(a) and Figure 1(b). Figure 1(c) shows a top-view optical image of the AP-HEMT, while Figure 1(d) provides a magnified view of the core active region, illustrating the spatial layout of the source, drain, and gate electrodes. The active channel area measures 25 μm ×10 μm (length × width) with $L_{GS}/L_{G}/L_{GD}$ = 3/5/17 μm. The AP-HEMT features $L_{AP}$ = 14 μm. Detailed geometric dimensions and epitaxial layer specifications are outlined in the Methods section and Table S1 (in the Supplementary Information). Cross-sectional transmission electron microscope (TEM) images in Figure 1(e) explicitly reveal the active passivation layer extending along the gate-to-drain region on the AlGaN surface. Notably, due to the non-uniformity of the etching process, the thickness of the active passivation layer gradually tapers along the gate-to-drain direction, eventually reaching the targeted dimension.

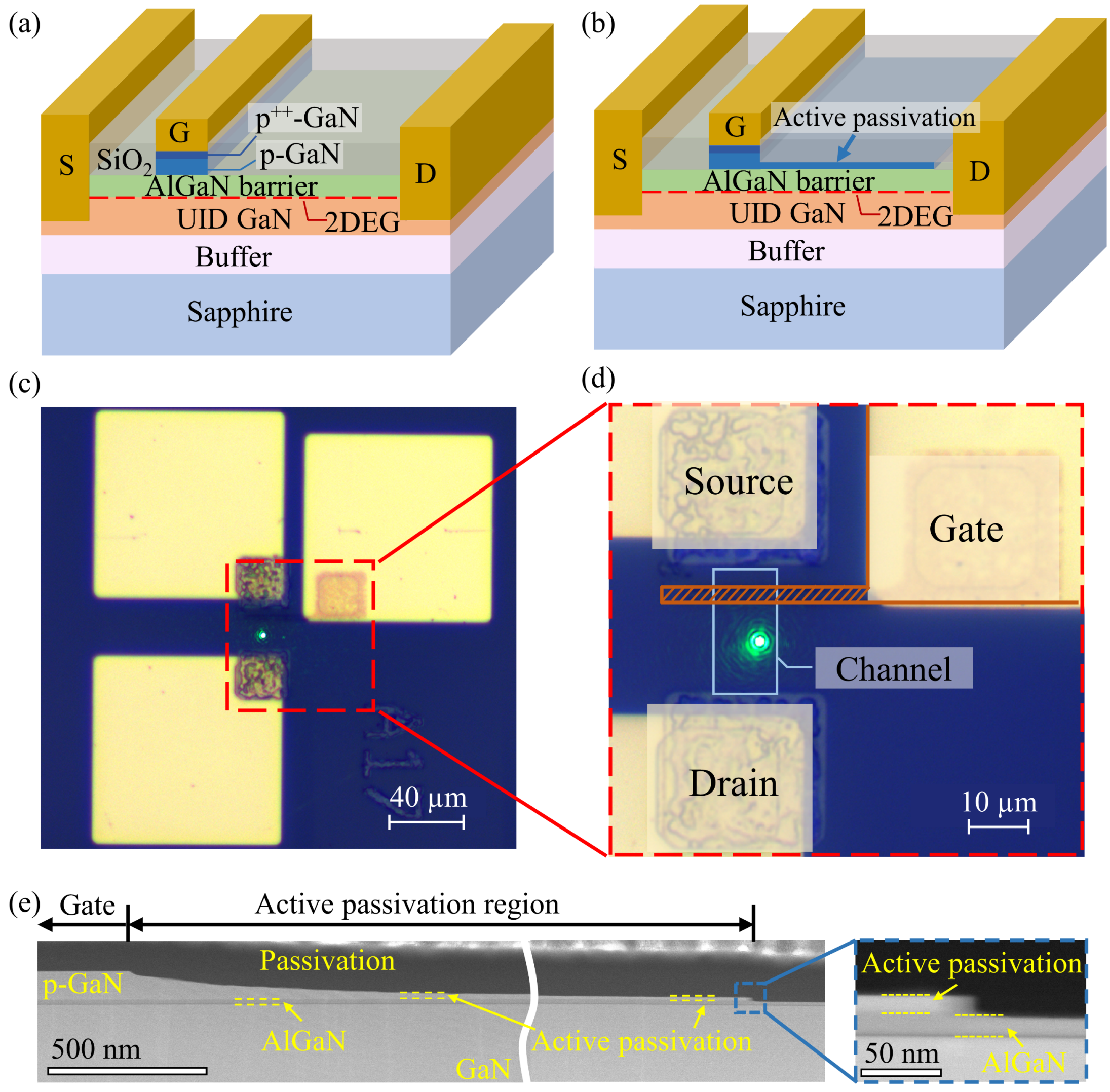

Figure 1. Sample structure. (a) Schematic structure of the Conv-HEMT and (b) the AP-HEMT. (c) Top-view optical image of the AP-HEMT, featuring an active channel area measuring 25 µm × 10 µm. $L_{GS}/L_G/L_{GD}$ = 3/5/17 µm for both devices. (d) Magnified view of the active channel region demonstrating the source, drain, and gate electrodes. (e) TEM images around the active passivation region of the AP-HEMT.

**Electric-field-driven heat-source modulation**

Before analyzing the thermal and mechanical profiles, it is essential to evaluate how the active passivation layer fundamentally modulates the electrical behaviors of the AP-HEMT. The current-voltage ($I$–$V$) properties are experimentally characterized, while the electric field and Joule heat density profiles are simulated using Sentaurus TCAD.

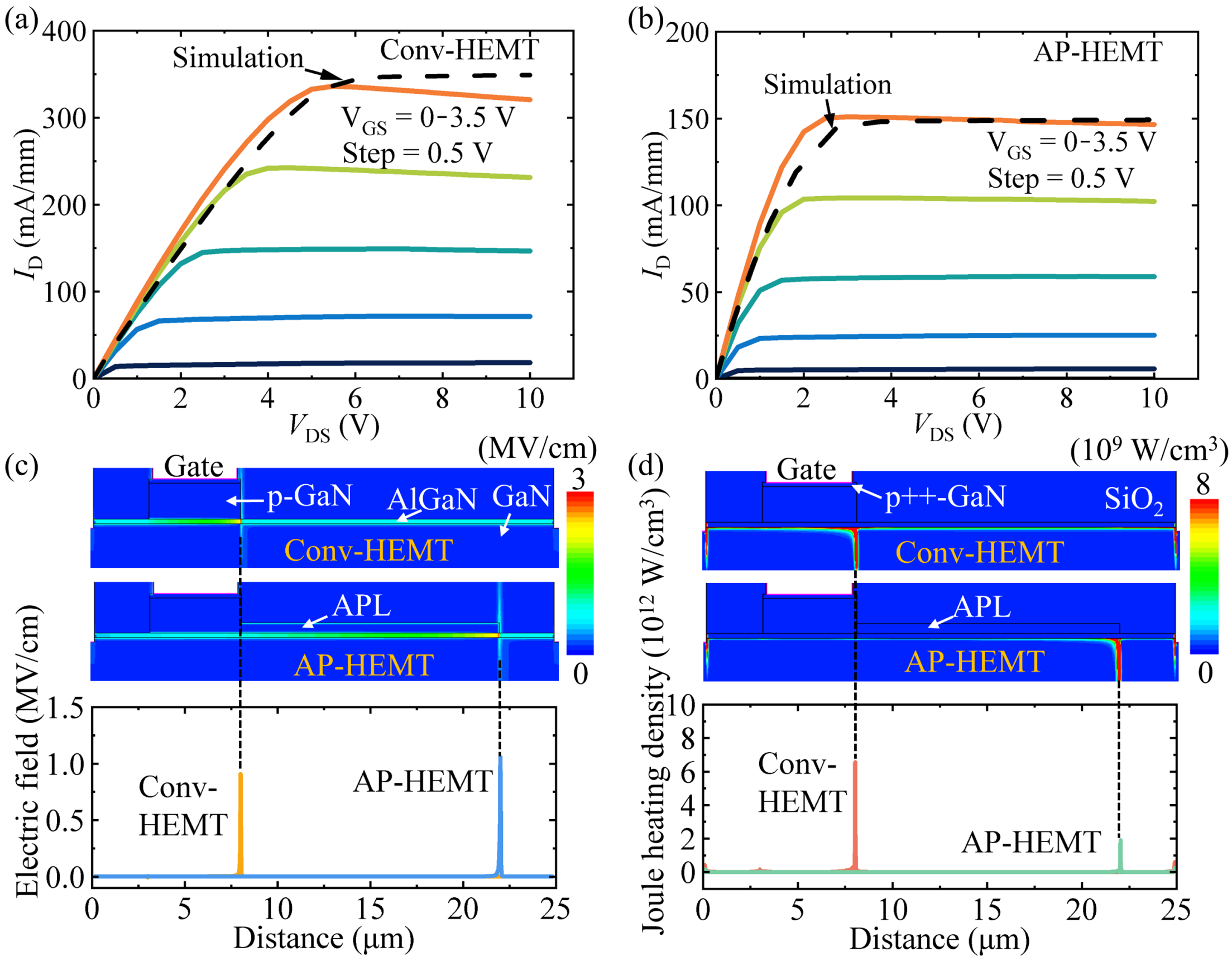

Figure 2. Measured and simulated electrical properties for the Conv-HEMT and the AP-HEMT with $L_{GS}/L_G/L_{GD}$ = 3/5/17 μm. The AP-HEMT features $L_{AP}$ = 14 μm. $I_D$–$V_{DS}$ output characteristics of (a) the Conv-HEMT and (b) the AP-HEMT. Solid lines represent the experimentally measured data, while the dashed lines represent the simulated data at $V_{GS}$ = 3.5 V. (c) 2D electric field simulations of the Conv-HEMT (top) and the AP-HEMT (middle), along with the extracted lateral electric field profile along the GaN surface for both devices (bottom). (d) Joule heating simulations of the Conv-HEMT (top) and the AP-HEMT (middle), along with the extracted Joule heating density

along the GaN surface for both devices (bottom). The simulated profiles in (c) and (d) are obtained under the same bias condition of $V_{DS}$=10 V and $V_{G}$=3.5 V.

Figures 2(a) and (b) show the output characteristics of the Conv-HEMT and AP-HEMT. At the same gate bias, the AP-HEMT exhibits a reduced saturation drain current and a lower saturation voltage compared to the Conv-HEMT. This is attributed to the active passivation layer shifting the channel pinch-off point from the gate edge to its drain-side end, where pinch-off occurs at a lower drain voltage. Notably, at $V_{G}$ = 3.5 V, the Conv-HEMT exhibits a prominent decrease in saturation current with increasing $V_{DS}$. This is a result of the temperature rise induced by Joule heating [28,29]. The simulated saturation current is higher than the experimental data because thermal feedback was omitted in the TCAD model. In contrast, the saturation current of the AP-HEMT remains nearly constant and aligns with the simulated data, indicating suppressed localized heat accumulation.

In Figure 2(c), the AP-HEMT exhibits a distinct spatial shift of the peak electric field toward the drain, with a comparable amplitude to the corresponding Conv-HEMT. Since the Joule heating density is equivalent to the product of the current density and the electric field, the primary heat source in the AP-HEMT is deterministically relocated to the drain-side terminus of the active passivation layer, as explicitly illustrated in Figure 2(d). Furthermore, due to the reduced $I_{D}$, the peak Joule heat density of the AP-HEMT is much lower compared to the Conv-HEMT.

**In-situ thermal-mechanical visualization**

Building upon the insights gained from the electric field and Joule heat redistribution, the steady-state thermal and mechanical behaviors of the GaN HEMTs under bias were systematically characterized to experimentally verify the efficacy of this proactive structural engineering in achieving multi-physics co-optimization. Utilizing non-destructive micro-Raman thermography based on the two-peak fitting method described in the Methods section, 2D high-resolution maps of the temperature rise and in-plane thermal stress were acquired for both Conv-HEMTs and AP-HEMTs under various power dissipation levels. The spatial mapping was performed with a step size of 1 μm along both the *x*-axis (along the channel length) and *y*-axis (along the channel width), and the measured area was 4 μm × 23 μm in the center of the active channel. Due to the partial optical obstruction of the channel by the gate and drain metallization, the accessible scanning length is shorter than the physical channel dimensions.

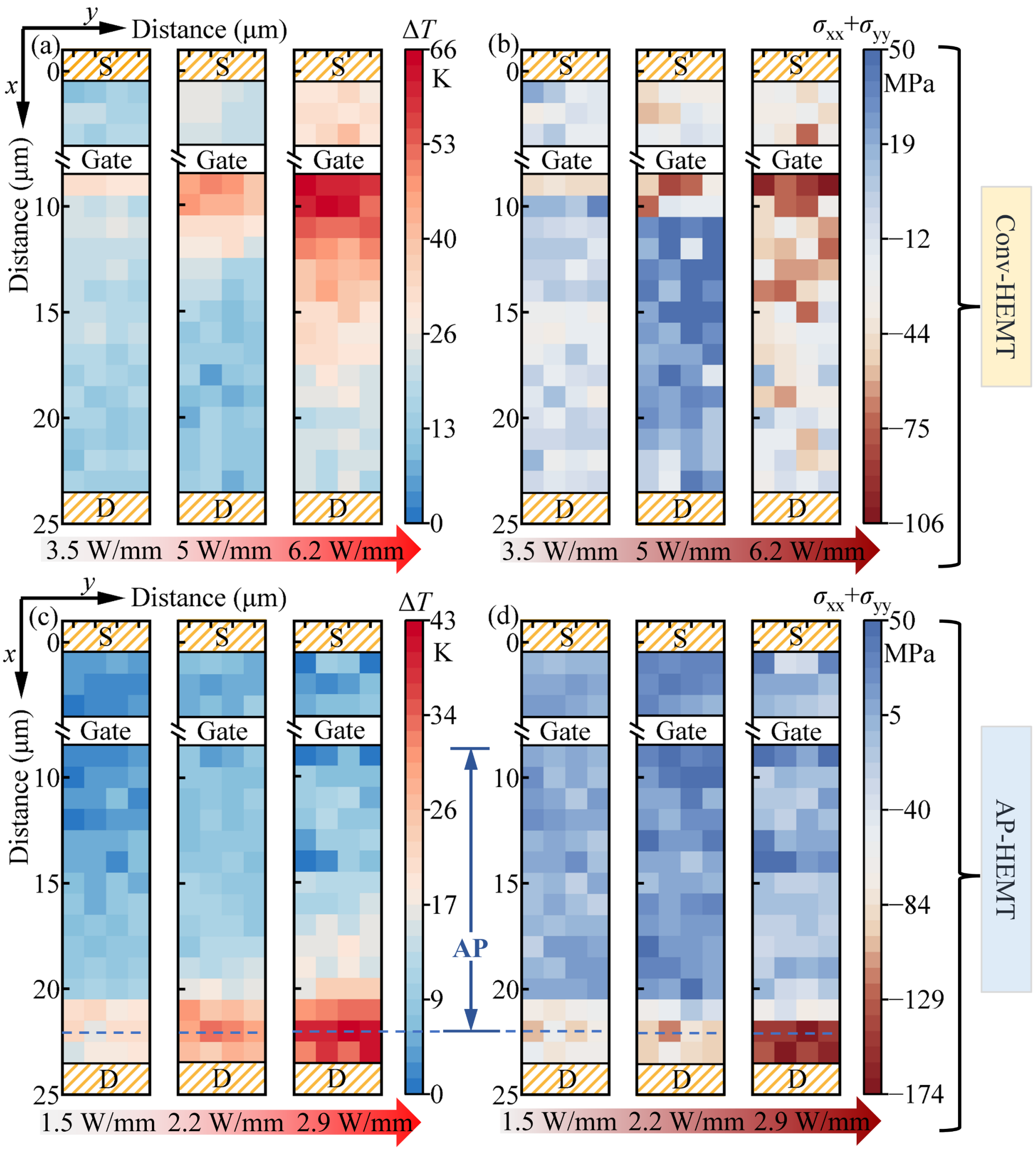

Figure 3. Measured thermomechanical profiles of the partially scanned active channel for the Conv-HEMT and the AP-HEMT with $L_{GS}/L_{G}/L_{GD}$ = 3/5/17 μm under various power dissipation levels. The AP-HEMT features $L_{AP}$ = 14 μm. (a) Temperature rise and (b) corresponding in-plane thermal stress distributions of the Conv-HEMT operating at 3.5 W/mm (10 V, 350 mA/mm), 5.03 W/mm (15 V, 335 mA/mm), and 6.24 W/mm (20 V, 312 mA/mm), respectively. (c) Temperature rise and (d) corresponding in-plane thermal stress distributions of the AP-HEMT operating at 1.5 W/mm (10 V,

150 mA/mm), 2.19 W/mm (15 V, 146 mA/mm), and 2.85 W/mm (20 V, 142.5 mA/mm), respectively. $\sigma_{xx}$ and $\sigma_{yy}$ are the biaxial in-plane stress components, with negative values representing compressive stress. Here, the *x*- and *y*- axes correspond to the channel length and width directions, respectively. The measurement window for each device was located as close to the center of the active region as possible.

Figure 3 illustrates the measured 2D distributions of the temperature rise ($\Delta T$) and the corresponding in-plane thermal stress ($\sigma_{xx} + \sigma_{yy}$) for the Conv-HEMT and the AP-HEMT under various power densities. As shown in Figure 3(a), a prominent hotspot emerges and progressively intensifies near the drain-side gate edge as the operating power scales up. Specifically, the peak temperatures rises under the three different power levels are recorded as 32 °C, 49 °C, and 65 °C, respectively. This thermal localization is a direct consequence of the prominent peak electric field at the drain-side edge of the gate contact, which triggers intense Joule heating within a confined channel region.

The resulting in-plane thermal stress profiles, shown in Figure 3(b), exhibit a strong spatial correlation with the temperature distributions, as both are derived from the same experimental dataset using the two-peak fitting method. Due to the localized thermal expansion constrained by the cooler surroundings, a distinct compressive stress region develops at the hotspot location. The spatial co-localization of the peak temperature and maximum compressive stress underscores the critical reliability issues of the Conv-

HEMT.

In contrast to the Conv-HEMT, the temperature profiles in Figure 3(c) clearly demonstrate the effective thermal modulation capability enabled by the active passivation layer. The localized hotspot is successfully driven away from the gate edge and relocated towards the drain contact at the end of the active passivation layer, consistent with the spatial shifts of the peak electric field and Joule heating density. The maximum temperature rises under the three different power levels are 22 °C, 34 °C, and 42 °C, respectively. The reduced temperature rise at the same bias voltage is primarily attributed to the lower saturation current, which suppresses the total Joule heating. This electro-thermal optimization is highly advantageous for achieving robust and high-reliability power switching applications. Unlike traditional field plates that form multiple electric field sub-peaks underneath while retaining a weakened peak at the gate edge[30,31], the AP technology enables a complete migration of the peak electric field away from the gate, thereby achieving a full hotspot relocation.

Figure 3(d) shows that the in-plane thermal stress profiles of the AP-HEMT closely mirror this spatial migration of the hotspot. A localized compressive stress zone develops at the shifted hotspot position near the drain side under the three different power levels. Therefore, the region beneath the gate is effectively shielded, experiencing mitigated thermal stress gradients. This dramatic relocation of both the peak temperature and compressive thermal stress underscores the efficacy of the AP technology in realizing the electro-thermal-mechanical co-optimization.

Notably, the aforementioned AP-HEMT with $L_{GS}/L_G/L_{GD}/L_{AP}$= 3/5/17/14 μm exhibits higher thermal stress than the Conv-HEMT, while possessing a lower temperature rise. To investigate the underlying mechanisms behind this behavior, we fabricated another set of Conv-HEMT and AP-HEMT with $L_{GS}/L_G/L_{GD}$= 3/5/7 μm, where the AP-HEMT features $L_{AP}$ = 4 μm. Under the same measurement conditions, 2D micro-Raman mapping was conducted to characterize the temperature rise ($\Delta T$) and the corresponding in-plane thermal stress ($\sigma_{xx} + \sigma_{yy}$) for both devices under different power levels, as illustrated in Figure 4. The corresponding electrical properties are illustrated in the Supplementary Information.

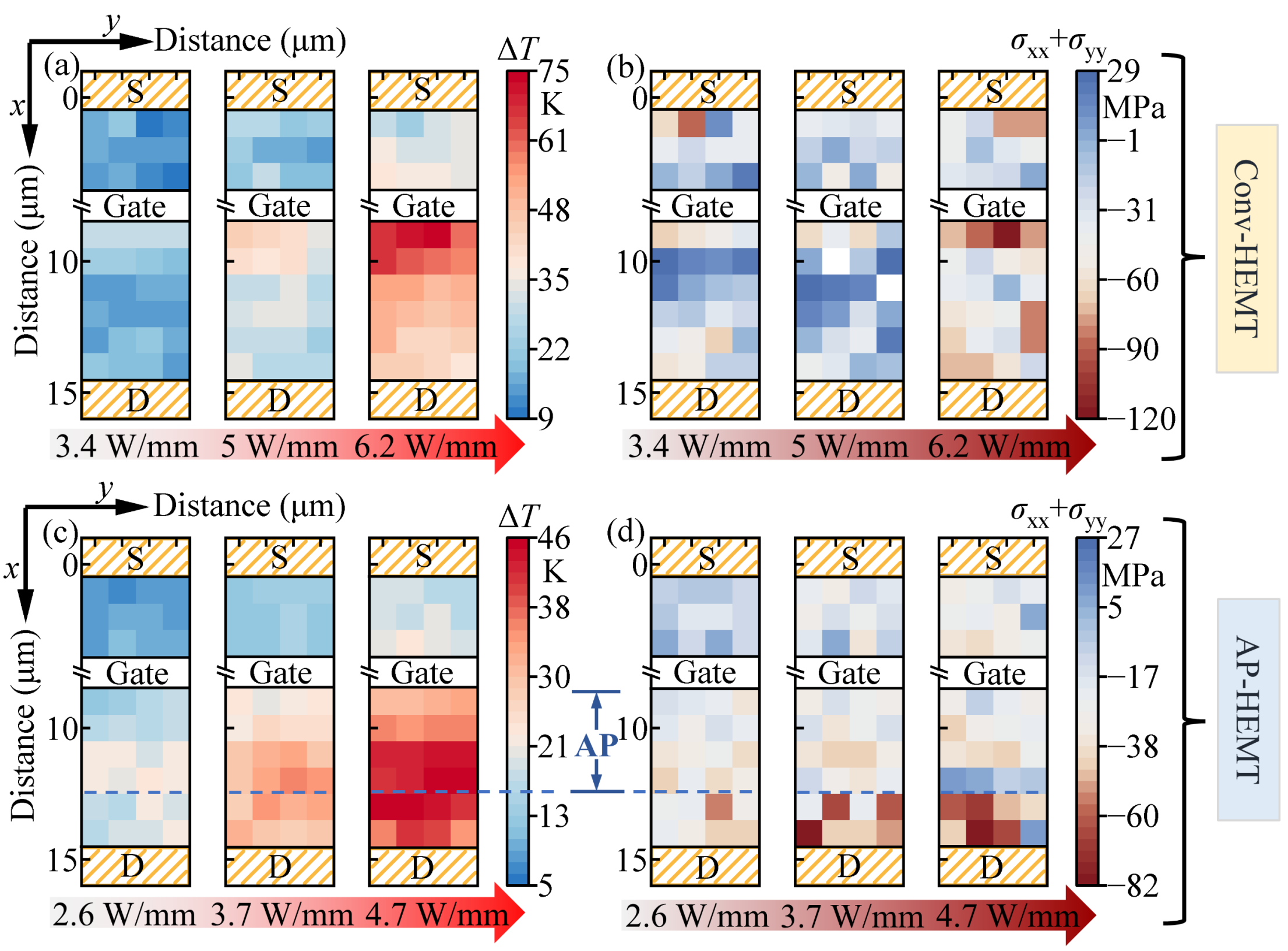

Figure 4. Measured thermomechanical profiles of the partially scanned active channel for the Conv-HEMT and the AP-HEMT with $L_{GS}/L_G/L_{GD}$ = 3/5/7 μm under various power dissipation levels. The AP-HEMT features $L_{AP}$ = 4 μm. (a) Temperature rise and (b) corresponding in-plane thermal stress distributions of the Conv-HEMT operating at 3.4 W/mm (10 V, 340 mA/mm), 5 W/mm (15 V, 330 mA/mm), and 6.2 W/mm (20 V, 310 mA/mm), respectively. (c) Temperature rise and (d) corresponding in-plane thermal stress distributions of the AP-HEMT operating at 2.6 W/mm (10 V, 260 mA/mm), 3.7 W/mm (15 V, 247 mA/mm), and 4.7 W/mm (20 V, 237 mA/mm), respectively. $\sigma_{xx}$ and $\sigma_{yy}$ are the biaxial in-plane stress components, with negative values representing compressive stress. Here, the *x*- and *y*- axes correspond to the channel length and width directions, respectively. The measurement window for each device was located as close to the center of the active region as possible.

From Figure 4, the AP-HEMT ($L_{GS}/L_G/L_{GD}$ = 3/5/7 μm) exhibits the similar peak temperature to the AP-HEMT ($L_{GS}/L_G/L_{GD}$ = 3/5/17 μm), but demonstrates a lower peak thermal stress. A direct spatial comparison between Figure 3(c) and Figure 4(c) indicates that the temperature gradient near the hotspot in the AP-HEMT ($L_{GS}/L_G/L_{GD}$ = 3/5/17 μm) is visibly steeper than that in the AP-HEMT ($L_{GS}/L_G/L_{GD}$ = 3/5/7 μm). This steeper temperature gradient is the primary driver for the elevated thermal stress. This behavior may be attributed to the evolution of the electric field profile. As $V_{DS}$ increases, the peak electric field rises initially, followed by a subsequent elevation in the field away from the peak. Under the same $V_{DS}$, a smaller $L_{GD}$ yields a comparable peak electric field, but leads to a higher field away from the peak due to the reduced channel length, as shown in Figure S2. This elevated field effectively smooths the spatial distribution of Joule heating, thereby resulting in a milder temperature gradient and mitigated thermal stress.

### $L_{AP}$-induced hotspot dynamics

A key advantage of active passivation technology is its capability to precisely regulate the spatial distribution of the internal heat source. To establish the quantitative correlation, we systematically evaluated a series of AP-HEMTs with varying $L_{AP}$.

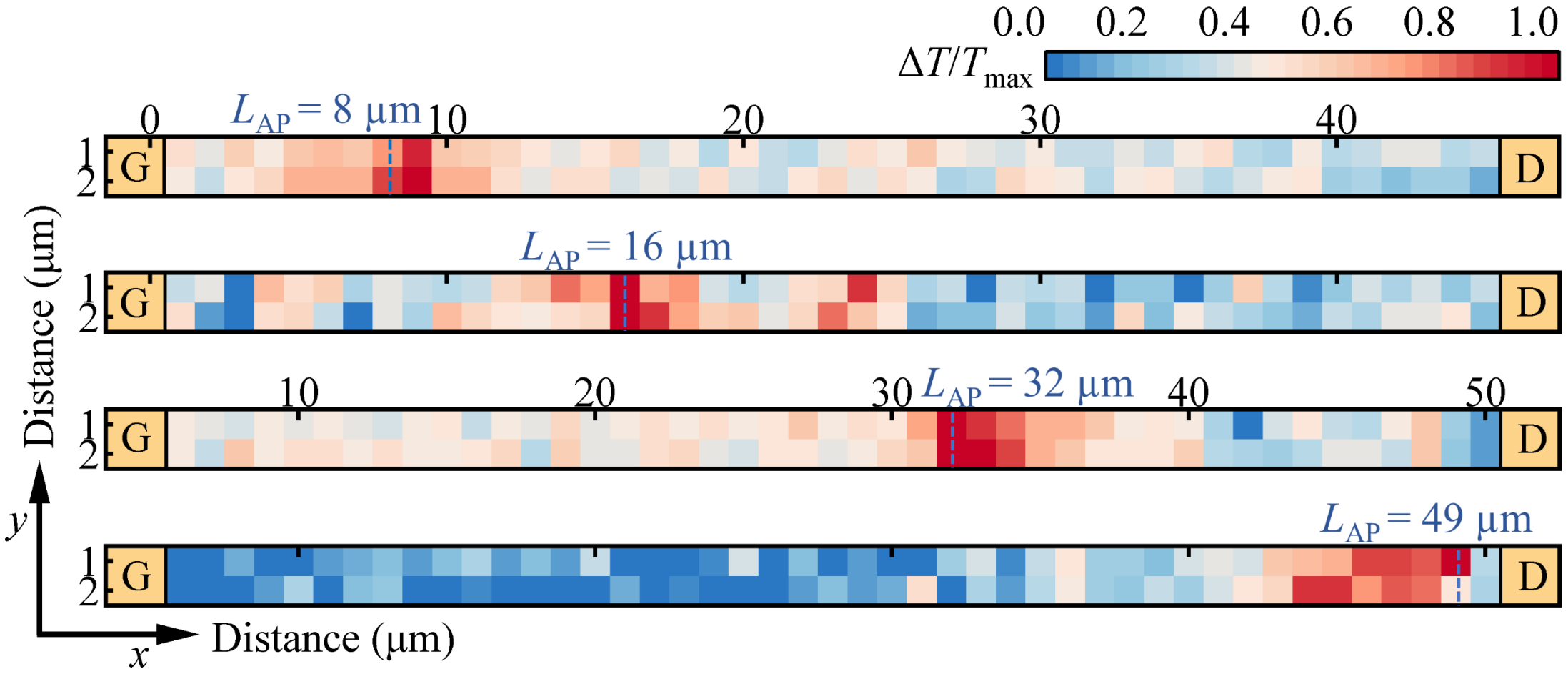

Figure 5. Normalized temperature rises ($\Delta T/T_{max}$) and hotspot positions of the AP-HEMTs ($L_{GS}/L_G/L_{GD}$ = 3/5/52 μm) with varying active passivation lengths of 8 μm, 16 μm, 32 μm, 49 μm, respectively. Along the channel length ($x$-axis), the measurement window spanned 1–45 μm (near the gate) for the former two AP-HEMTs and 6-50 μm (near the drain) for the latter two devices. Note that the electrode blocks ('G' and 'D') serve as schematic markers for device orientation rather than exact physical positions.

As shown in Figure 5, a distinct spatial migration of the heat source is observed as $L_{AP}$ scales up. By precisely tracking the coordinates of the peak temperature across AP-HEMTs with varying $L_{AP}$, we demonstrate a one-to-one spatial correlation between the hotspot position and the active passivation length, validating the peak electric field modulation mechanism. Because the channel pinch-off point and the corresponding peak electric field are pinned to the drain-side terminus of the active passivation layer, the peak Joule heating zone follows this variation dynamically. Consequently, manipulating $L_{AP}$ provides a robust strategy to redistribute the internal electric field, thermal profiles, and mechanical stress within the critical gate-to-drain access region.

**Thermal-mechanical simulation**

To cross-validate the experimental micro-Raman results and provide theoretical insights for the structural optimization, a coupled heat transfer and solid mechanics multi-physics model of the AP-HEMT ($L_{GS}/L_{G}/L_{GD}/L_{AP}$ = 3/5/17/14 μm) was built, and subsequent finite-element simulations were carried out using COMSOL Multiphysics. Specifically, the heat source was configured based on the shifted peak Joule heating profile extracted from the TCAD simulation shown in Figure 2(d). The simulated temperature rise and thermal stress profiles were compared against the measured data.

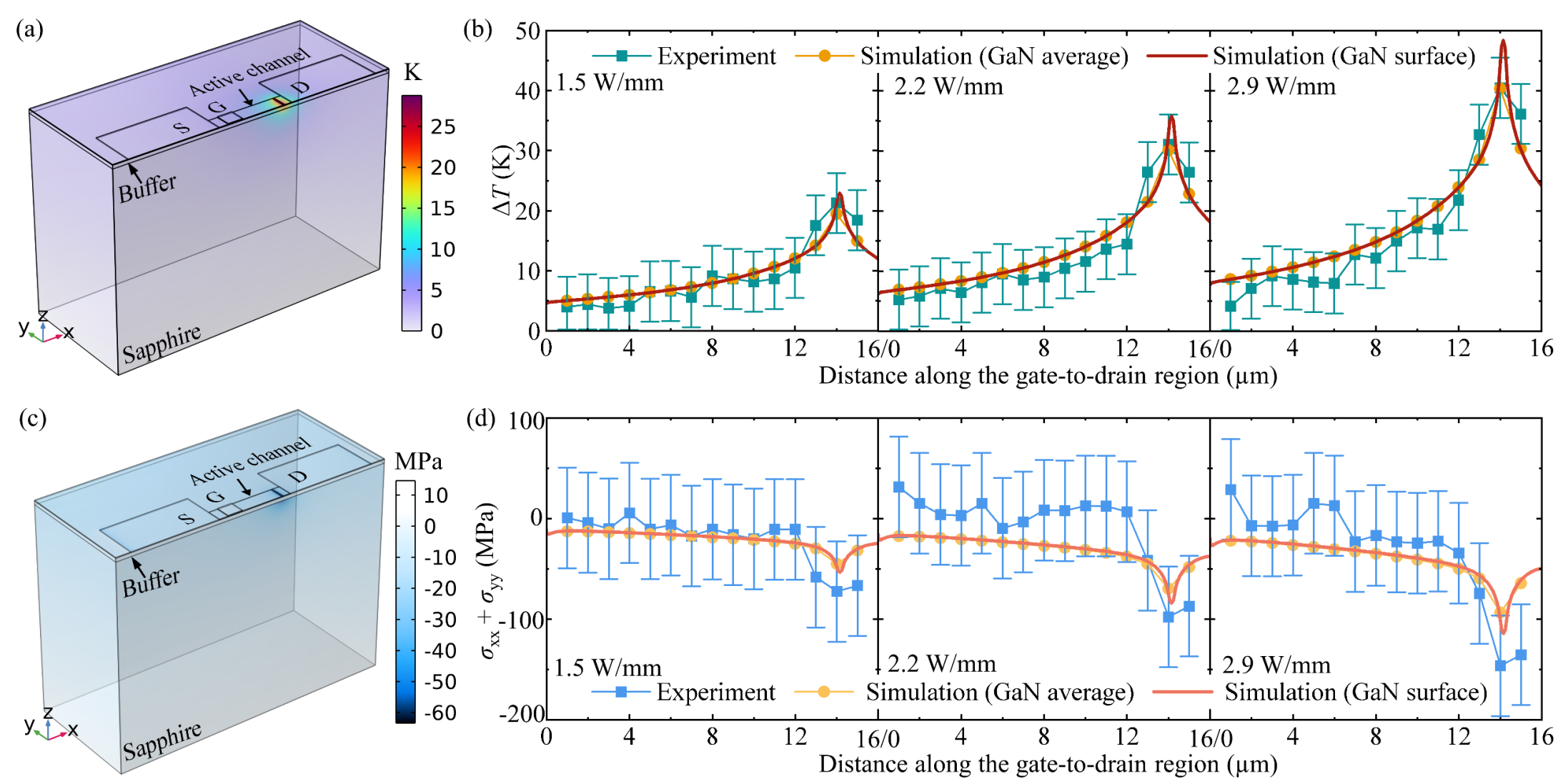

Figure 6. Thermal and mechanical simulations of the AP-HEMT with $L_{GS}/L_{G}/L_{GD}/L_{AP}$ = 3/5/17/14 μm. (a) Simulated 3D temperature rise distribution of the AP-HEMT operating at 1.5 W/mm. (b) Comparison between the measured and simulated temperature rises along the gate-to-drain region under different power densities of 1.5 W/mm, 2.2 W/mm, and 2.9 W/mm, respectively. (c) Simulated 3D thermal stress distribution at 1.5 W/mm. (d) Comparison between the measured and simulated in-plane thermal stress under the corresponding power densities. Squares indicate the experimental results with error bars (±5 K for temperature rise and ±60 MPa for thermal stress). Dots denote the thickness-averaged simulated values in the GaN layer, while solid lines represent the simulated GaN surface profiles along the center of the active channel. $\sigma_{xx}$ and $\sigma_{yy}$ are the biaxial in-plane stress components.

As illustrated in Figure 6(a), the simulated 3D temperature rise distribution reveals a highly localized thermal concentration at the drain-side end of the active passivation layer. This severe temperature gradient corresponds to the peak Joule heating region, which causes localized compressive thermal stress shown in figure 6(c). Figure 6(b) displays the lateral temperature profiles of the AP-HEMT ($L_{GS}/L_G/L_{GD}/L_{AP}$ = 3/5/17/14 μm) under different power densities. For a direct comparison with micro-Raman thermography, the simulated temperature rise is weighted by the Gaussian laser profile and averaged across the GaN layer according to Equation (1) [32,33]:

$$T_{\mathrm{m}} = \frac{\int_0^{\infty} T(r)\exp(-\frac{r^2}{r_0^2})rdr}{\int_0^{\infty} \exp(-\frac{r^2}{r_0^2})rdr} \tag{1}$$

where $T(r)$ is the local simulated temperature rise; $r$ is the radial distance from the center of the laser spot and $r_0$ is the characteristic Gaussian laser radius. The thickness-averaged simulated results show excellent agreement with the experimental data within the ±5 K error margin under three different power densities. The thermal simulation provides a reference for designing the maximum operating power of GaN HEMTs.

As shown in Figure 6(d), the in-plane thermal stress ($\sigma_{xx} + \sigma_{yy}$) induced by thermal expansion mismatches and temperature gradients is evaluated. Corresponding to the thermal hotspot shift, the peak compressive stress similarly shifts toward the active passivation layer edge. The thickness-averaged simulated stress in the GaN layer agrees well with the measured values within the error margin. The difference between the two data comes from the uncertainty of stress coefficients used in the two-peak fitting method. The elastic stress along the $z$-axis is zero based on the assumption that the GaN layer can freely move along the $z$-direction.

## Conclusions

In summary, we have established a systematic electro-thermal-mechanical co-optimization and design framework for GaN transistors based on AP technology. By actively modulating internal electric field distributions, the AP-HEMT successfully relocates the hotspot away from the gate edge to the drain-side end of the active passivation layer. In-situ micro-Raman thermography experimentally confirms a strict one-to-one spatial correlation between the hotspot position and $L_{AP}$. This hotspot migration effectively reshapes the temperature and thermal stress profiles. Notably, the short-channel AP-HEMT ($L_{GD}$ = 7 μm) exhibits lower peak thermal stress than the long-channel device ($L_{GD}$ = 17 μm), as the smoother electric field profile mitigates localized temperature gradients. By providing direct, visualized experimental evidence for designable electric field modulation and heat-source engineering via AP technology, this study offers a new perspective on proactive device thermal management through tuning hotspot locations. This provides a vital design reference for the next-generation high-power and high-reliability electronic devices.

## Methods

### *Fabrication process*

The epitaxial stack consists of a 15-nm $Al_{0.2}Ga_{0.8}N$ barrier layer and a 200-nm unintentionally doped GaN (UID GaN) channel layer grown on a 430-μm sapphire substrate via a 1.2-μm buffer layer. The p-GaN gate region consists of a 76.6-nm p-GaN layer (magnesium-doped at $3 \times 10^{19}$ $cm^{-3}$) capped with an 11.5-nm $p^{++}$-GaN contact layer (doped at $3 \times 10^{20}$ cm$^{-3}$). A 140-nm $SiO_2$ layer serves as the passivation layer. Regarding the device layout, the active channel width is 10 μm, and the source/drain ohmic contact pads measure 30 μm × 30 μm. For devices with $L_{AP}$ of 4 μm and 14 μm, the corresponding $L_{GD}$ are 7 μm and 17 μm, respectively. For $L_{AP}$ = 8, 16, 32, and 49 μm, $L_{GD}$ is fixed at 52 μm. The gate-to-source distance ($L_{GS}$ = 3 μm) and gate length ($L_G$ = 5 μm) remain constant across all devices. Ti/Al/Ni/Au metallization was utilized for the source/drain ohmic contacts while Ni/Au stacks were employed to form the ohmic gate contacts. Further details on the fabrication process are available in reference [22]. Structural and layout details are summarized in Table S1.

***Electrical simulation***

In the TCAD model, the geometric dimensions of the simulated devices were set according to the fabricated devices. Both devices feature $L_{GS}/L_G/L_{GD}$ = 3/5/17 μm, with an additional $L_{AP}$ of 14 μm for the AP-HEMT. Spontaneous and piezoelectric polarizations at the AlGaN/GaN heterojunction were used in the TCAD simulation. The main physical models implemented in the simulations include doping-dependent mobility, field-dependent mobility, Shockley-Read-Hall (SRH) recombination, and the polarization model [34–36].

***Raman thermography***

Raman thermography, with a sub-micron spatial resolution, is widely used to measure the active channel temperature of GaN HEMTs. As the temperature rises, the frequencies of $E_2$(high) and $A_1$(LO) phonons redshift due to anharmonic effects [37]. In theory, either Raman peak frequency can be used to determine the temperature rise in operating GaN HEMTs [14]. However, along with the temperature rise, converse piezoelectric and thermoelastic stresses also develop, causing a blueshift in phonon frequencies. If the effects of stress are not taken into account, the measured temperature will be underestimated [10]. Converse piezoelectric stress is generated by the electric field perpendicular to the channel surface, and its effect on Raman thermography can be eliminated by setting the pinch-off state as the reference state [38]. Thermoelastic stress arises from a temperature gradient. The two-peak fitting method [39] was employed to decouple and quantify the temperature rise and thermal stress. This method does not require a calibration process, thereby reducing measurement time and avoiding the introduction of additional measurement errors.

The Raman spectra of the Conv-HEMTs and AP-HEMTs were measured using a confocal micro-Raman system (Nanobase XperRam S) in a backscattering configuration. The laser spot diameter was 1 μm, as measured by the knife-edge method (in Supplementary Information) using a 50× objective with a numerical aperture (NA) of 0.5. A 532-nm sub-bandgap laser was utilized for excitation at a power of 18 mW, where laser-induced heating and its effect on the device current were negligible. Using an 1800 lines/mm grating, the spectra were collected with a resolution of 0.15 $cm^{-1}$ over the range of 350–850 $cm^{-1}$, as shown in Figure S3. The $A_{1g}$ mode of the sapphire substrate, and the $E_2$(high) and $A_1$(LO) modes of GaN were detected and well-fitted using Lorentz functions. Subsequently, the active channel was mapped in the same configuration as the above single-point measurement, with a step size of 1 μm both the x-axis and y-axis.

**Thermal-mechanical simulation**

To verify the measured data and investigate the factors affecting the temperature distribution, a heat transfer finite-element model was established using COMSOL Multiphysics. The AP-HEMT was constructed based on the dimensions detailed in Table S1. The substrate size was set large enough to be considered semi-infinite, ensuring that further increases in substrate size resulted in no change in the temperature rise [40]. To improve computational efficiency, a half-model with a symmetric boundary condition was used in the simulation. The 15-nm AlGaN barrier layer was omitted due to its minimal effect on temperature [39], while the source, drain, and gate metal electrodes were retained, as they expand the heat spreading surface and may have a significant effect on the temperature distribution [41,42]. Because the temperature rise is relatively small (less than 50 °C for the AP-HEMT operating at 2.85 mW/mm), the temperature dependence of material properties was neglected. The thermal properties of AlGaN, GaN, AlN, and sapphire are listed in Table S2. In addition, the thermal boundary conductance (TBC) values of the GaN/AlN and AlN/sapphire interfaces were set to 500 $MW/m^2$-K [43,44] and 160 $MW/m^2$-K [45], respectively. Since the $p^{++}$-GaN layer is only 17 nm thick, the $p^{++}$-GaN and p-GaN layers were modeled as a single layer, with a effective thermal conductivity of 50 W/m-K, based on the references [46,47]. its accuracy has a negligible effect on the total temperature rise. The bottom and sides of the device were assumed to be isothermal surfaces at 20 ℃, while the top surface was subjected to a natural convection boundary condition with a heat transfer coefficient of 10 $W/m^2$-K.

Corresponding to the peak electric field, A heat source with dimensions of 0.2 × 0.02 × 10 μm was implemented at the top surface of the GaN layer beneath the APL end [48]. The dissipation power was the same as in the experiment. To resolve steep thermal gradients, an ultra-fine tetrahedral mesh was generated within the active region, especially around the localized heat source, to accurately capture peak temperatures. Farther away from this critical zone, the mesh element size gradually increases to save computational time. Finally, the governing Fourier heat transfer equation was solved to obtain the temperature distribution.

Based on the temperature rise calculated by the heat transfer model, a solid mechanics module was built and coupled via thermal expansion multi-physics. Omitting the thin AlGaN layer also has a negligible effect on the simulated thermal stress of the underlying, much thicker GaN layer [39]. Based on the symmetry structure, the top surface was set as free, while the bottom and sides were subjected to fixed constraints. The mechanical properties of the key materials are listed in Table S2. The p-GaN layer was assumed to have the same mechanical properties as GaN. Notably, the anisotropy of thermal expansion coefficients was considered. Finally, by solving linear elastic equations, the thermoelastic stress was obtained.

## CRediT authorship contribution statement

Yicheng Wei: Writing – original draft, Writing – review & editing, Investigation, Data curation, Formal analysis, Visualization. Sihang Liu: Writing – original draft, Writing – review & editing, Resources, Data curation, Formal analysis, Visualization. Zimu Jiang: Data curation. jinquan Zhang: Software. Zifeng Huang: Writing – review & editing. Han Yang: Data curation. Yang He: Writing – review & editing. Jin Wei: Writing – review & editing, Conceptualization, Methodology, Validation, Supervision, Funding acquisition. Zhe Cheng: Writing – review & editing, Conceptualization, Methodology, Validation, Supervision, Funding acquisition.

## Declaration of competing interest

The authors declare that they have no known competing financial interests or personal relationships that could have appeared to influence the work reported in this paper.

## Acknowledgments

This study was supported by the National Key Research and Development Program of China (Grant No. 2024YFA1207901), the National Natural Science Foundation of China (Grant No. 62574007, T2550270).

## Data availability

The data that support the findings of this study are available from the corresponding authors upon reasonable request.